\documentclass[aps,prl,twocolumn,groupedaddress,floatfix]{revtex4-2}

\usepackage{graphicx}
\usepackage{natbib,hyperref}
\usepackage[english]{babel}
\usepackage{color}
\usepackage{float}

\begin{document}

	\title{Nanoscale Skyrmions on a Square Atomic Lattice}
	
	\author{Reiner Br\"uning*}
	\author{Andr\'e Kubetzka}
	\author{Kirsten von Bergmann}
	\author{Elena Y. Vedmedenko}
	\author{Roland Wiesendanger}
	
	\affiliation{Department of Physics, University of Hamburg, Jungiusstr.~11, 20355 Hamburg, Germany}
	
	\date{\today}
	
	\begin{abstract}
		Spin-polarized scanning tunneling microscopy has been applied to study non-collinear spin textures of a Mn monolayer on a four-fold symmetric W(001) substrate revealing a zero-field spin spiral ground state and two different types of rotational domain walls. With an applied magnetic field of 9 T, we observe a coexistence of the spin spiral and the skyrmion phase, even though a previous theoretical study reported that the phase transition occurs at 18 T. The skyrmions show a roughly hexagonal arrangement despite the square lattice symmetry of the W(001) substrate. Based on a reduced set of energy parameters, we are able to describe the experimental findings and to analyze the topological properties of the rotational domain walls.
	\end{abstract}
	
	\maketitle
	
	Since the first experimental observation of skyrmion lattices in chiral magnets \cite{muhlbauerS2009,yuN2010} and ultrathin magnetic films \cite{heinzeNP2011} as well as the detection and controlled manipulation of individual skyrmions \cite{rommingS2013}, skyrmion-based devices have become promising candidates for advancing the field of spintronics based on the unique properties of skyrmions, in particular their particle-like nature, their nano-scale size, their enhanced stability due to a non-trivial real-space topology, and the low current densities needed for their transport \cite{sampaioNN2013,nagaosaNN2013,wiesendangerNRM2016,fertNRM2017,everschor-sitteJoAP2018}. Epitaxially grown ultrathin film systems are particularly suited to investigate the fundamental properties of skyrmions \cite{rommingPRL2015,Leonov_2016}, e.g. the influence of the underlying crystal lattice symmetry, the response to electric or magnetic fields, as well as skyrmion creation and annihilation processes. While earlier investigations of skyrmions in ultrathin films have exclusively concentrated on three-fold symmetric crystal lattices (e.g. Fe on Ir(111) \cite{heinzeNP2011,hsuNC2018}, Pd/Fe on Ir(111) \cite{rommingS2013,rommingPRL2015}, Rh/Co on Ir(111) \cite{meyerNC2019}, Co on Ru(0001) \cite{herveNC2018}), and already revealed the influence of the atomic layer stacking on skyrmionic states \cite{vonbergmannNL2015}, the impact of a square crystal lattice symmetry on the properties of non-collinear spin textures, including skyrmions, has not been experimentally addressed so far.
	
	Here we study a monolayer (ML) of Mn on a four-fold symmetric W(001) substrate by using spin-polarized scanning tunneling microscopy (SP-STM) with atomic-scale spatial resolution \cite{wiesendangerRMP2009}. A spin spiral ground state in zero field is found experimentally with two different types of rotational domain walls which are adequately described by atomistic spin dynamics simulations. Earlier combined experimental and theoretical investigations of this system showed that the magnetic ground state is a left-handed cycloidal spin spiral with two degenerate rotational domains and a period of about 2.2~nm \cite{ferrianiPRL2008}. However, in contrast to previous theoretical predictions of a phase transition from the spiral to a skyrmion phase occurring at a magnetic field of about 18~T \cite{nandyPhys.Rev.Lett.2016}, we find that individual skyrmions can already be observed at significantly lower fields of about 9~T. By fitting a minimal model to the density functional theory (DFT) data \cite{ferrianiPRL2008} with a reduced number of parameters we achieve a good description of our experimental findings.
	
	\begin{figure}[htbp]
		\includegraphics[width=1\columnwidth]{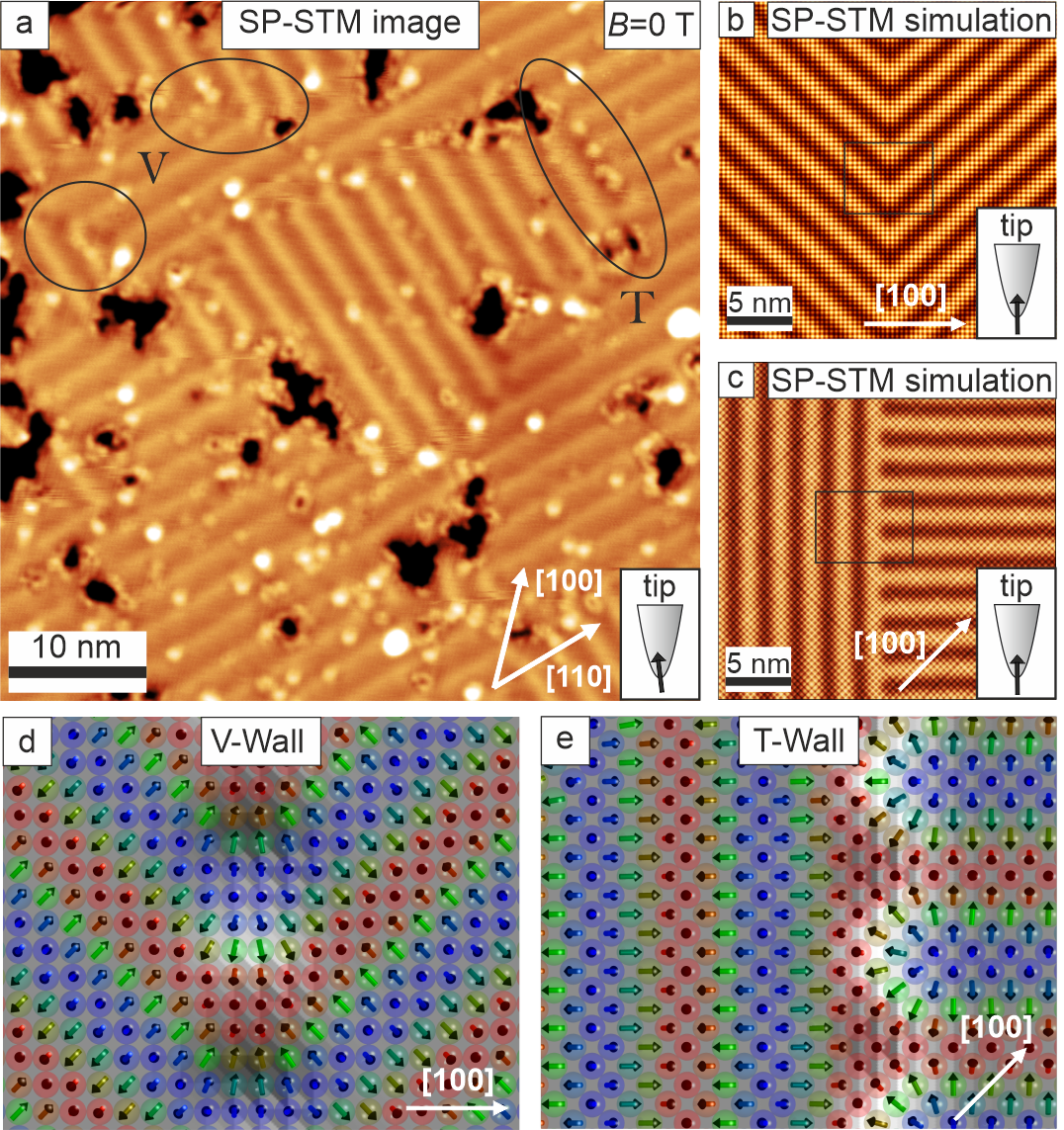}
		\caption{(a) Spin-resolved constant-current STM image of 0.9 ML Mn/W(001) at $B = 0$ T in remanence obtained with a magnetic tip mostly sensitive to the out-of-plane component of the sample magnetization ($U = -0.05$ V, $I = 3$ nA). The ellipses mark two different types of rotational domain wall configurations of the spin spiral phase. (b),(c) SP-STM simulations of a V-wall and a T-wall, respectively. The walls were set up and then relaxed with an atomistic spin dynamics code (see Supplemental material S4,S5 for further details). (d),(e) Sketches of the spin configurations of both types of rotational domain walls within the areas marked in (b),(c). Each arrow represents the magnetic moment of a single Mn atom, green indicates in-plane moments, while red and blue indicate opposite out-of-plane moments.}
		\label{fig1}
	\end{figure}
	
	The samples were prepared by growing pseudomorphic Mn MLs on an atomically clean W(001) substrate under ultra-high vacuum conditions. Mn was evaporated from a Knudsen cell with typical deposition rates of approximately 0.2~ML/min (see Supplemental material S1 for further details). The SP-STM measurements were performed at 4.2 K using a home-built low-temperature STM instrument equipped with a superconducting magnet providing an out-of-plane field of up to 9~T. A Cr bulk tip was used for spin-sensitive STM studies \cite{wiesendangerRMP2009}.
	
	Figure 1(a) shows a constant-current SP-STM image of the Mn ML on W(001) obtained at zero field. The stripes running along $<$110$>$ directions result from the spin spiral ground state. The period of the spin spiral is 2.2 nm, i.e. about five times the diagonal of the square atomic lattice. This is in agreement with previous experimental results \cite{ferrianiPRL2008}. Based on DFT calculations, the origin of this spin spiral was attributed to frustrated exchange interactions together with a significant Dzyaloshinskii-Moriya interaction (DMI) of strength $D = 4.6$ ~meV/atom and an out-of-plane magnetocrystalline anisotropy of $K_z = 3.6$~meV/atom \cite{ferrianiPRL2008}. The wave vector of this spin spiral is strongly coupled to the four-fold symmetric atomic lattice resulting in the formation of two rotational domains. A comparably small spin spiral domain is visible in the center of the SP-STM image in Fig.~1(a). We find that the size of the domains is governed by the number of defects in the ultrathin Mn film, i.e. Mn clusters of varying sizes (bright spots in Fig.~1(a)) as well as Mn vacancy islands (dark spots in Fig.~1(a)). In addition we observe a larger number of domain boundaries in the remanent state, as in Fig.~1(a), compared to the virgin state (see Supplementary S2 for the virgin state). As a consequence of the formation of rotational spin spiral domains, we observe two different types of domain walls, i.e. V- and T-walls, which are indicated by black ellipses in Fig.~1(a).
	
	For understanding the formation and atomic-scale properties of the experimentally observed magnetic textures we have considered a Hamiltonian describing a system of interacting spins up to third-nearest neighbours, including DMI as well as magnetocrystalline anisotropy (see Supplementary, S4). We were able to reproduce the spin spiral periodicity using a pair-wise isotropic Heisenberg exchange of $J_1 = +18.7$~meV/atom, $J_2 = -4.4$~meV/atom and $J_3 = -2.65$~meV/atom between nearest-, next-nearest, and next-next-nearest neighbours, respectively. In the following, we use this parameter set to investigate the atomic-scale properties of the magnetic textures as well as their field dependence by solving the Landau-Lifschitz-Gilbert equation.
	
	Fig.~1(b) shows the result of an SP-STM contrast simulation of a simulated V-wall, characterized by mirror-symmetry of the adjacent spiral domains. The directions of the magnetic moments within the black rectangle of Fig.~1(b) are sketched in Fig.~1(d). The spin spiral period is about 4.5 times the diagonal of the square atomic lattice, in good agreement with the experimental results. A close inspection of the transition region between the two rotational domains reveals that lines of parallel magnetic moments show a rounded V-shaped domain wall. The width of the wall is on the order of a few lattice constants, and the energy cost of this V-wall with respect to a perfect spin spiral is 12.7~meV/nm. While the spin structure is coplanar within the domains, the spins form a truly three-dimensional magnetic texture within the wall, meaning that topological charges (Q) can emerge \cite{schoenherrNaturePhysics2018}. Indeed, we find a non-vanishing local topological charge density, indicated by the gray-scale shading in Fig.~1(d). However, adjacent local topological charges along the domain wall exactly cancel, making this wall topologically trivial.
	
	A corresponding simulation of the SP-STM image contrast and the spin structure of a T-wall is displayed in Figs.~1(c) and 1(e), respectively. Here, the path of the wall is parallel to one of the adjacent spin spiral wave vectors, and perpendicular to the other one. The energy of the T-wall is comparable to that of the V-wall, being 15.7~meV/nm. However, in contrast to the V-wall, the T-wall carries a net topological charge: each cap of the spin-down lines (blue in Fig.~1(e)), coming from the right, represents half of a skyrmion, and consequently this results in a net topological charge of 0.5 per spin spiral period within the T-wall. This analysis shows that V-walls and T-walls represent topologically distinct defect lines in the spin texture, the former without topological charge, and the latter with $Q =$ 0.5 per spin spiral period.
	
	\begin{figure*}[t]
		\includegraphics[width=0.78\textwidth]{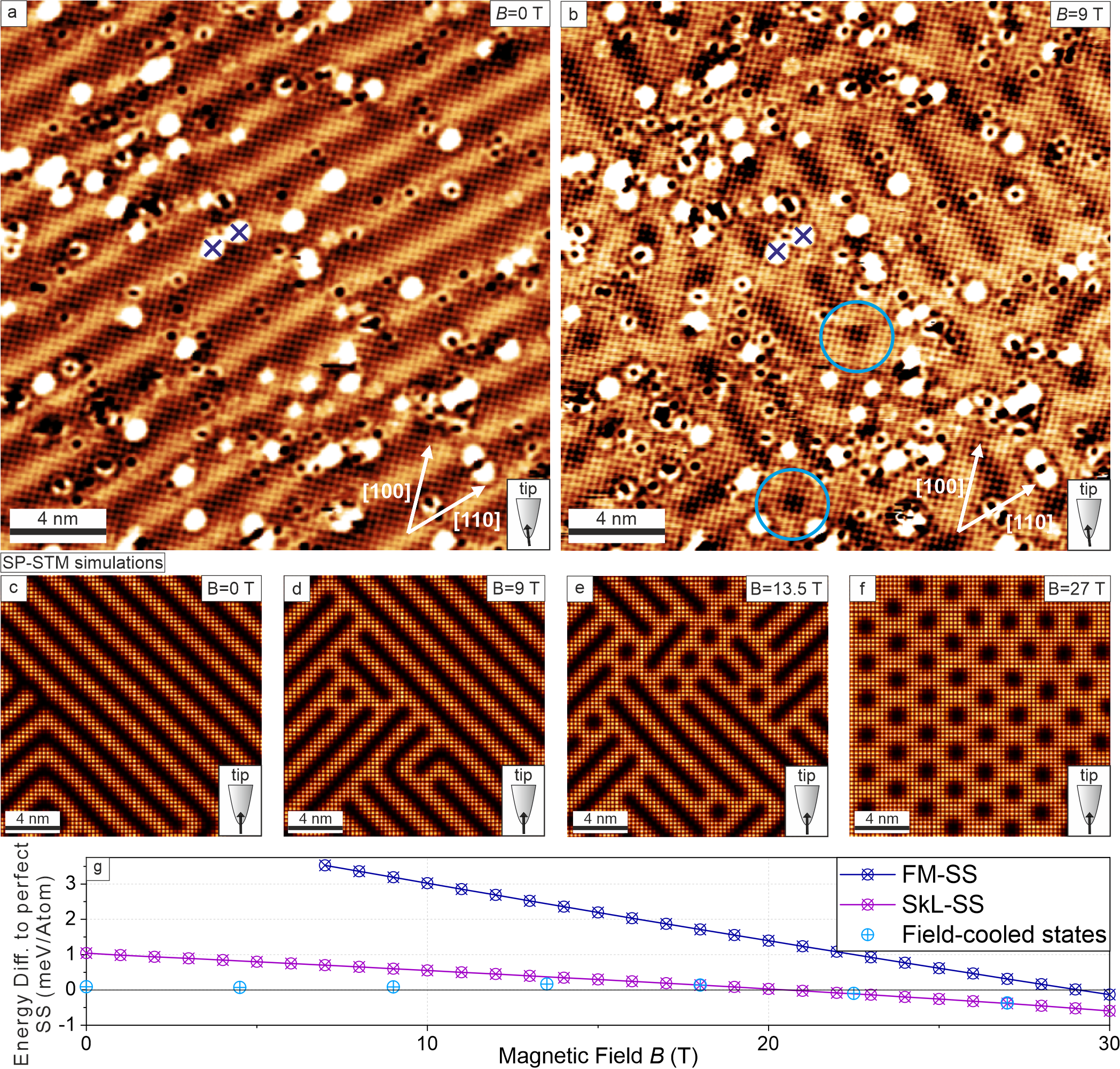}
		\caption{(a) SP-STM constant-current image of a single-domain spin spiral state in the Mn ML on W(001) at $B = 0$~T. (b) Image of the same area at $B = 9$~T showing a mixed state of small spiral segments and skyrmions (both (a) and (b): $U = -0.05$~V, $I = 3$~nA). (c)-(f) SP-STM simulations of magnetic states obtained using field-cooling at different $B$-fields as indicated. (g) Energy of the skyrmion lattice (SkL) and the ferromagnetic (FM) state relative to the perfect spin spiral (SS) state at different magnetic fields as obtained from simulations using field-cooling. Blue markers indicate the energies of the magnetic states displayed in (c)-(f) as well as in the Supplement S6.}
		\label{fig2}
	\end{figure*}
	
	When an external magnetic field is applied perpendicularly to the magnetic film we observe that the virgin zero-field state of a single spin spiral domain breaks up into numerous stripe segments of different length and orientation, see Figs.~2(a),(b); note that this is the same sample area, see clusters marked by crosses. The spin texture also exhibits small individual objects (two are marked by blue circles in Fig.~2(b)) that we identify as magnetic skyrmions as shown in the following. Previous DFT-based Monte Carlo simulations for 1 ML Mn/W(001) have predicted the transition from the spin spiral to the skyrmion lattice phase to occur at about 18~T \cite{nandyPhys.Rev.Lett.2016}, and also the parameter set used in our work leads to a phase transition at 20~T, see purple line in Fig.~2(g). At $B = 9$~T the energy difference between the spin spiral and skyrmion lattice phase is $-0.6$~meV/atom.
	
	To obtain an understanding why already at 9~T a mixed phase of spin spirals and skyrmions is observed in our experiments, we performed additional simulations with the stated parameter set for different external magnetic field strengths, cf. Figs.~2(c)-(g). At zero field, cf. Fig.~2(c), we find a spin spiral strictly coupled to the atomic lattice, with some V- and T-walls, in good agreement with the experimental findings presented in Fig.~1(a). For simulations with applied magnetic fields we use a field-cooling procedure (for details see Supplemental material S4), and indeed we find spin textures with both spin spiral fragments and skyrmions, see Figs.~2(d),(e), as in the experiments. The slightly larger number of skyrmions in the experimental data at 9~T may be attributed to the presence of defects. Additionally, the parameter set might be improved, e.g. by including higher-order exchange interactions \cite{heinzeNP2011}. Comparison of the energies of the simulated states (blue markers in Fig.~2(g)) with those of the perfect spin spiral state shows that they are very close to each other. The field-cooled state at $B = 13.5$~T is a mixed phase containing several skyrmions, but it has an energy penalty with respect to the ground state spin spiral of only 0.17~meV/atom. At 27~T a field-cooled simulation leads to a pure skyrmion lattice phase.
	
	\begin{figure}[htbp]
		\includegraphics[width=1\columnwidth]{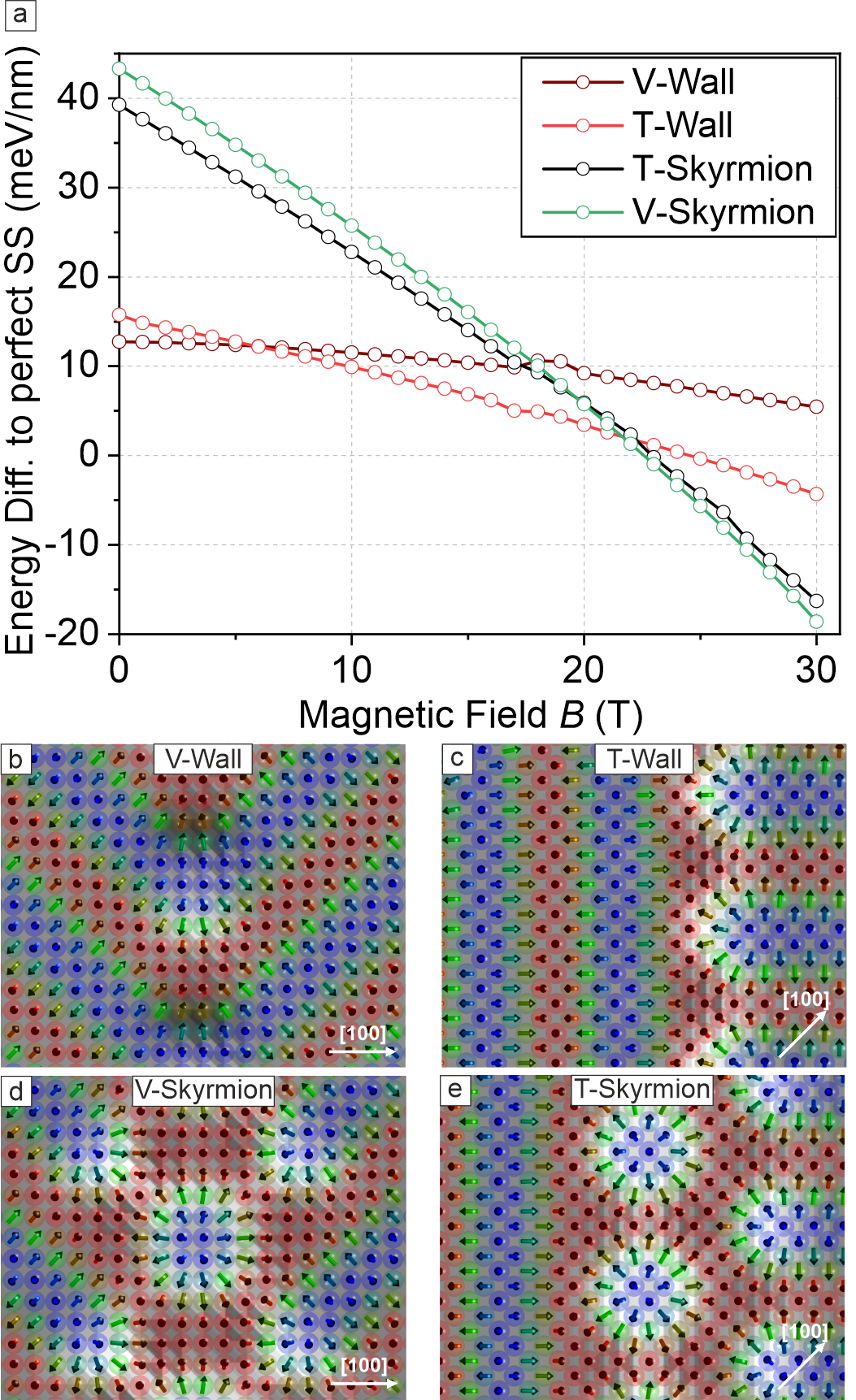}
		\caption{Magnetic field dependence of the energy of different types of rotational domain walls. (a) Energy per nm relative to the spin spiral for each type of rotational domain wall as presented in (b)-(e) at different magnetic fields. (b),(d) Spin structures of V- and T-wall. (d),(e) Spin structures of domain walls with lines of skyrmions in their center. }
		\label{fig3}
	\end{figure}
	
	Based on our model for the different types of rotational domain walls as presented in Fig.~1, we investigate the magnetic field dependence of the wall energies, see Fig.~3(a). Since we observe magnetic skyrmions at a magnetic field of 9~T (see Fig.~2), an extension of the model for the rotational domain walls (Figs.~3(b) and 3(c)) is required. This has been done by adding skyrmions in the center of each type of domain wall (see Figs.~3(d) and 3(e)). The resulting spin textures can indeed be found as building blocks of the spatially inhomogeneous magnetic states in Fig.~2. In agreement with the experimental data, the energy cost for introducing rotational domain walls is lowered with increasing magnetic field.
	
	\begin{figure}[htbp]
		\includegraphics[width=1\columnwidth]{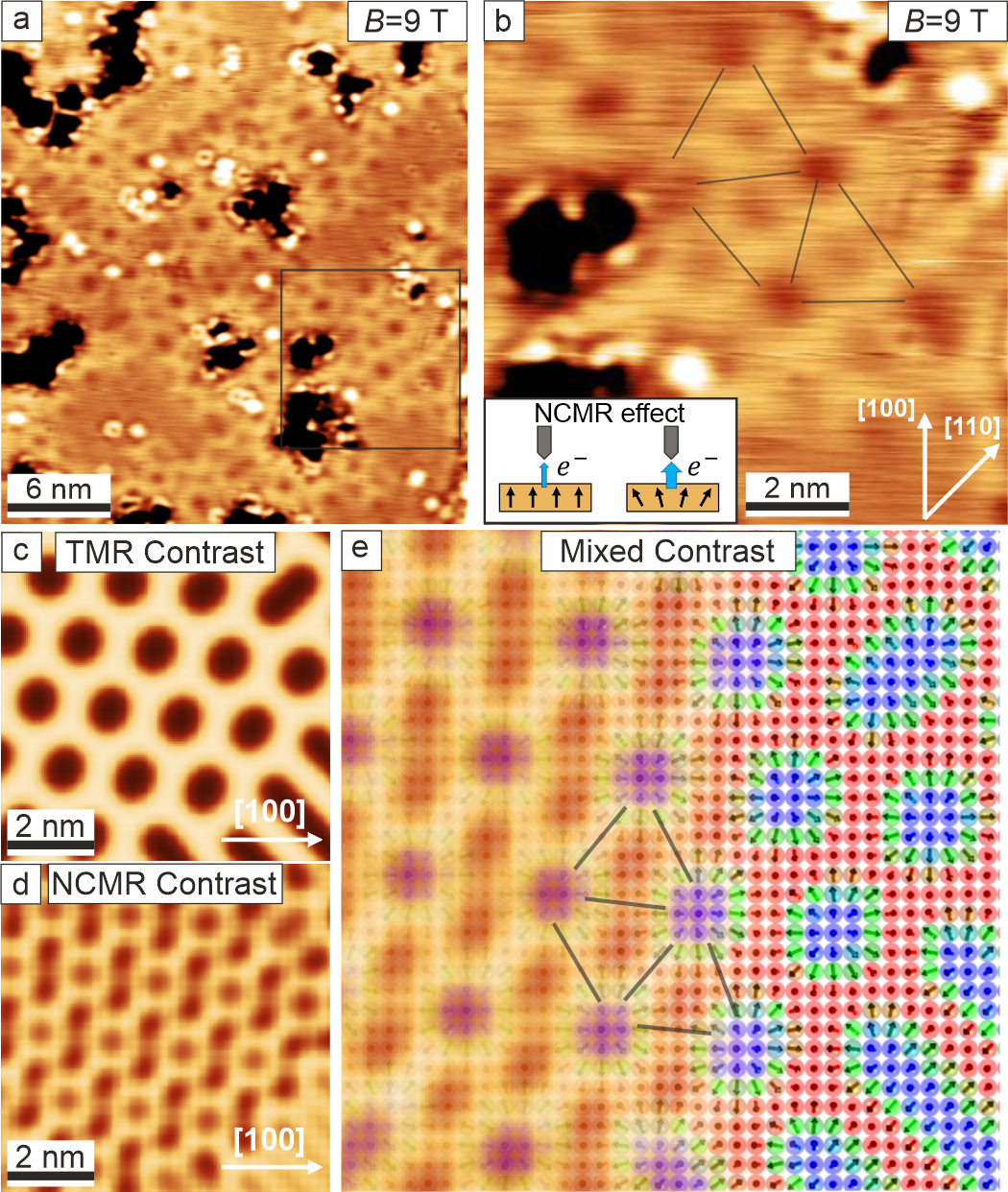}
		\caption{(a) Constant-current STM image of the Mn ML on W(001) measured at $B = 9$~T and at a comparably small bias voltage of $U = 0.01$~V. (b) Closer view of the sample area indicated by the black box in (a). This region shows a roughly hexagonal pattern of dark dots ($U = 0.01$~V, $I = 1$~nA). (c),(d) STM simulations using only TMR and only NCMR contrast, respectively, of a magnetic state obtained by field-cooling at 18~T. (e) STM simulation of the same magnetic state using both TMR and NCMR with a ratio of 1:45. The right-hand side shows a sketch of the spin configuration.}
		\label{fig4}
	\end{figure}
	
	The constant-current STM image displayed in Fig.~4(a) was obtained at a comparably small bias voltage of $U = +10$~mV. In this low-bias regime, the observed STM image contrast changes. In several parts of the image a pattern characterized by dark dots is visible. The distance between these dots is roughly 1.3-1.8 nm, i.e. smaller than the typical skyrmion-skyrmion distance of 2.5~nm (see Supplement S3). Close inspection reveals that some dots are slightly larger and darker, see the zoom-in image of Fig.~4(b), and arrange in a roughly hexagonal pattern with additional less dark dots around them. The distance between the larger dots is about 2.5~nm, i.e. the expected skyrmion-skyrmion distance. If the larger dots indicate skyrmion positions, then the question arises what the origin of the small dots is. From previous STM measurements of magnetic skyrmions it is known that in addition to the tunnel magnetoresistance (TMR) effect \cite{wiesendangerRMP2009} also the non-collinear magnetoresistance (NCMR) effect can contribute to the tunnel current \cite{hannekenNN2015,crumNC2015,kubetzkaPRB2017}. It arises due to spin-mixing effects modifying the local electronic states depending on the details of the local spin texture. This results in a contrast between locally collinear and non-collinear spin configurations, see inset of Fig.~4(b). Simulations of skyrmion lattice states reveal that for larger skyrmion-skyrmion distances, NCMR contrast arises only at the positions of skyrmions. In contrast, for small skyrmion-skyrmion distances as in Mn/W(001), the NCMR contribution to the STM image leads to an additional signature in-between skyrmions (compare Figs.~4(c) and 4(d) for TMR and NCMR contrast of the same skyrmion lattice area). To reproduce the experimental observations of Fig.~4(a), we consider both, TMR and NCMR contrast contributions, see Fig.~4(e), where on the right-hand side also the underlying spin texture of the simulations is sketched. The purely NCMR-related dots arise between the skyrmions because there is a strong spatial variation of non-collinearity due to the small distance between the skyrmions.
	
	In summary, we have investigated two types of zero-field rotational domain walls with trivial and non-trivial topology and a mixed phase of spin spirals and skyrmions on a square atomic lattice. We were able to gain insight into the atomic-scale structure of these highly complex non-collinear spin textures by a combination of atomic-resolution SP-STM experiments and atomistic spin dynamics simulations. Magnetic field dependent investigations revealed a broad transition regime from the spin spiral to a skyrmionic phase starting below 9~T. This could be reproduced by using spin dynamics simulations.
	
	\begin{acknowledgments}
		R.B. and R.W. acknowledge financial support by the European Union via the ERC Advanced Grant ADMIRE (grant No. 786020). K.v.B. and A.K. acknowledge funding by the Deutsche Forschungsgemeinschaft (DFG, German Research Foundation) - 402843438; 408119516, 418425860.
	\end{acknowledgments}
	
	\bibliography{lab13}
	
\end{document}